\documentclass[sigconf]{acmart}

\usepackage{multirow}
\usepackage{ulem} 
\usepackage{enumitem}
\usepackage{graphicx}
\usepackage{subcaption}
\usepackage{amsmath,amsfonts}
\usepackage[utf8]{inputenc}
\usepackage{todonotes}
\usepackage{xcolor}
\usepackage{balance}
\usepackage{siunitx}

\usepackage[skip=2.5pt]{caption}
\usepackage[compact]{titlesec}

\newcommand{\ie}{\textit{i.e.}}
\newcommand{\etc}{\textit{etc.}}

\newcommand{\smar}{SMAR}
\newcommand{\header}[1]{{\flushleft \textbf{#1}}}
\newcommand\blfootnote[1]{%
  \begingroup
  \renewcommand\thefootnote{}\footnote{#1}%
  \addtocounter{footnote}{-1}%
  \endgroup
}

\AtBeginDocument{%
  \providecommand\BibTeX{{%
    \normalfont B\kern-0.5em{\scshape i\kern-0.25em b}\kern-0.8em\TeX}}}

\setcopyright{rightsretained}
\setcopyright{acmlicensed}
\copyrightyear{2023}
\acmYear{2023}
\acmDOI{10.1145/3539618.3591863}
\acmConference[SIGIR ’23]{the 46th International ACM SIGIR Conference on Research and Development in Information Retrieval}{July 23–27, 2023}{Taipei, Taiwan}
\acmPrice{15.00}
\acmISBN{978-1-4503-9408-6/23/07}

\begin{document}
\fancyhead{} 

\title{Semantic-enhanced Modality-asymmetric Retrieval for Online E-commerce Search}
\author{Zhigong Zhou$^\dagger\,$, Ning Ding$^\dagger\,$, Xiaochuan Fan, Yue Shang, Yiming Qiu, Jingwei Zhuo \\Zhiwei Ge, Songlin Wang, Lin Liu, Sulong Xu and Han Zhang$^*\,$}
\affiliation{
  \institution{JD.com, Beijing, China}
  \country{}
  \{\small zhouzhigong1, dingning36, xiaochuan.fan, yue.shang, qiuyiming3, zhuojingwei1, gezhiwei, wangsonglin3, liulin1, xusulong, zhanghan33\}@jd.com
}

\begin{abstract}
Semantic retrieval, which retrieves semantically matched items given a textual query, has been an essential component to enhance system effectiveness in e-commerce search.
In this paper, we study the multimodal retrieval problem, where the visual information (e.g, image) of item is leveraged as supplementary of textual infomration to enrich item representation and further improve retrieval performance.
Though learning from cross-modality data has been studied extensively in tasks such as visual question answering or media summarization, multimodal retrieval remains a non-trivial and unsolved problem  especially in the asymmetric scenario where the query is unimodal while the item is multimodal.
In this paper, we propose a novel model named {\smar}, which stands for \textbf{S}emantic-enhanced \textbf{M}odality-\textbf{A}symmetric \textbf{R}etrieval, to tackle the problem of modality fusion and alignment in this kind of asymmetric scenorio.
Extensive experimental results on an industrial dataset show that the proposed model outperforms baseline models significantly in retrieval accuracy. 
We have open source our industrial dataset~\footnote{https://github.com/jdcomsearch/jd-multimodal-data} for the sake of reproducibility and future research works.

\blfootnote{$^\dagger\,$ Both authors contribute equally}
\blfootnote{$^*\,$ corresponding author}
\end{abstract}


\maketitle

\section{Introduction}
Product search, as the most important service in an e-commerce system, provides the interface for users to get what they need from the returned items relevant to their input queries, thus contributes to the largest percentage of transactions among all channels~\cite{sondhi2018taxonomy,sorokina2016amazon}.
Modern search engine is usually designed as multi-stage pipelines which consist of query processing, candidate retrieval and ranking. In this paper, we focus on the candidate retrieval stage which aims to retrieve relevant products from billions of items collection.

\begin{figure}[t!]
\centering
\includegraphics[width=2.5in]{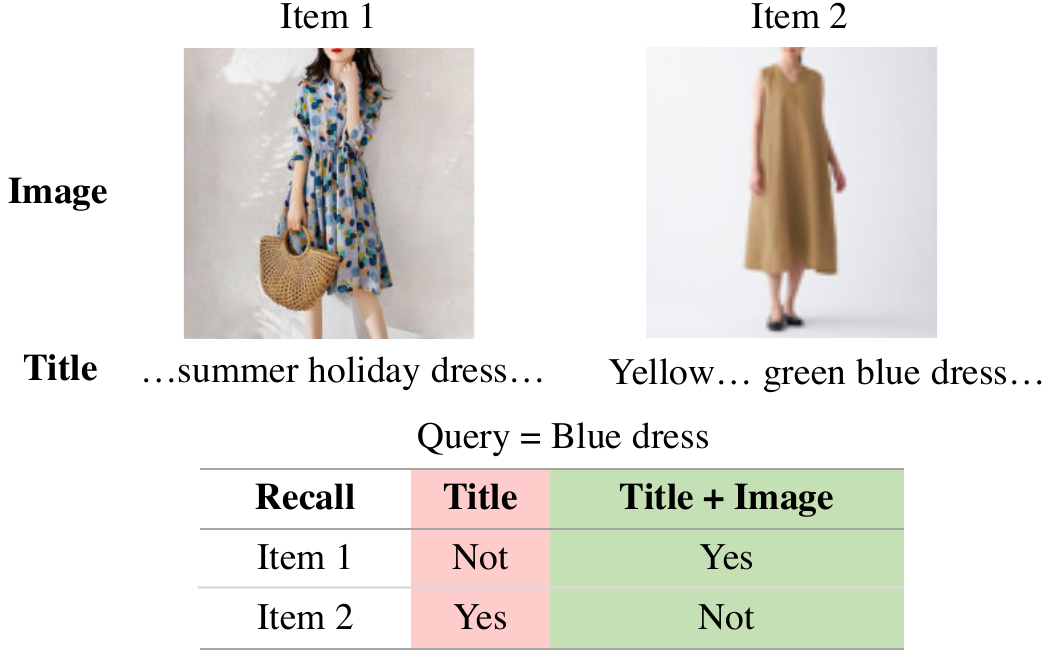}
\caption{Illustration of the scenarios in which image improves item retrieval.}
\label{fig:intro}
\vspace{-5mm}
\end{figure}

Benefiting from the development of Neural Language Processing technologies, most semantic retrieval models~\cite{huang2013learning,zhang2020towards,li2021embedding} achieve satisfactory performance by learning query and item representations with text features. While in e-commerce system, in addition to text features, there are many other resources with supplementary information that can be used to enrich the latent representation of items, such as images and videos. For example, as shown in Figure~\ref{fig:intro}, the red column are two bad cases in semantic model because of the attribute shortage or redundancy, while the green column are the correct result of multimodal model since the image supplies extra color information.
Thus how to import image information properly to improve the performance of semantic retrieval becomes an interesting and challenging problem.

Multimodal learning, which takes information from both text modality and vision modality, has been studied in many tasks, like visual question answering~\cite{wu2017visual,chen2022rethinking}, media summarization~\cite{hossain2019comprehensive,hu2022scaling}, image matching~\cite{gao2020fashionbert,jiang2021review},  multimodal retrieval~\cite{li2020oscar,lin2020interbert,lu2019vilbert,yu2021ernie}{\etc} 
However, to the best of our knowledge, most works have focused on the symmetric multimodal problems where two sides are represented by only one modality. Few works have dedicated to the asymmetric ones where two sides are represented by different number of modalities. In this paper, we aim to solve this kind of asymmetric retrieval problem, where the query is represented by text modality, the item is represented by both text and image modalities. The asymmetric retrieval problem faces two critical challenges: (1) how to improve the modality fusion to take supplementary but not redundant information from image. (2) How to solve the problem of asymmetric modality alignment.

In this paper, we propose a novel asymmetric multimodal model {\smar} to tackle the above challenges. The contributions of our work can be summarized as follow.
\begin{itemize}
    \item In Section~\ref{sec:method}, we propose a novel model which dedicates to the fusion of multiple modalities, the alignment of asymmetric modalities, and the contributions of different modalities.
    \item In Section~\ref{sec:exp}, we conduct extensive offline experiments and online A/B tests to demonstrate the effectiveness of our proposed model.
    \item We open source an industrial dataset which facilitates the reproducibility of our work and offers valuable research material for public works.
\end{itemize}

\section{Method}
\label{sec:method}

\begin{figure*}[th!]
\centering
\includegraphics[width=7in,height=2.3in]{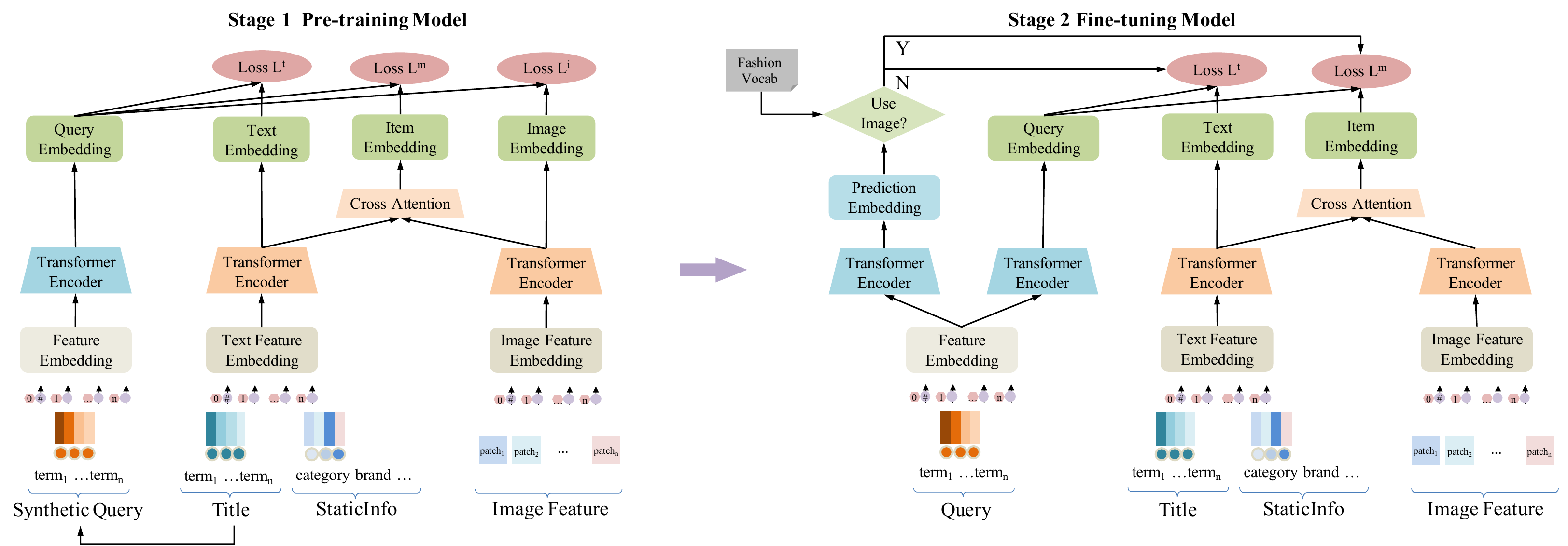}
\caption{Overview of the proposed model {\smar}.}
\label{fig:three_loss}
\end{figure*}

In this section, we first present some preliminaries: the standard two tower embedding model, which is the basis of many semantic retrieval models, as well as our work. Then we introduce the details of {\smar}, which consists of two stage models: a pre-training model and a fine-tuning model, as illustrated in Fig.~\ref{fig:three_loss}. 

\subsection{Preliminaries of Standard Embedding Retrieval Model}
A standard embedding retrieval model has a two tower architecture, which consists of a query tower Q and an item tower S. Given a query $q$ and an item $s$, we compute the score as
\begin{equation}
    f(q, s) = F(Q(q), S(s)),
\end{equation}
where $Q(q) \in \mathbb{R}^{k}$ is the output $k$-dimensional embedding of query tower, $S(s) \in \mathbb{R}^{k}$ is the output embedding of item tower with the same dimension, and $f(q, s)$ is the output of scoring function $F$ which calculates the similarity between query and item.
The design of two tower architecture is mainly due to the huge amount of candidate items in retrieval task.


\subsection{Text and Image Modality Representations}
Transformer~\cite{vaswani2017attention}, which has been widely used in text encoding, also shows outstanding performance in visual representation. In our work, we adopt transformer to the text and image representations, as well as the fusion of these two modalities. Following the two tower architecture, {\smar} contains the following 
 four towers.

\textbf{Query Tower} is constructed by three layers, which are feature embedding layer, transformer encoder layer and normalization layer. We can get the output of query tower as
\begin{align}
    H^q &= transformer^q(X^q), \\
    h^q &= cls(H^q), \\
    Q(q) &=norm(h^q),
\end{align}
where $X^q$ is the feature embedding of the textual query, $transformer^q$ is a $o^q$-heads, $l^q$-layers transformer encoder, $cls$ means to fetch the CLS token embedding of the transformer encoder output $H^q$ that is similar to BERT~\cite{devlin2018bert}, and $norm$ means the operation of L2 normalization. Here use $o^q=12$ and $l^q=4$.

\textbf{Item Text Tower} is designed to encode text features of item, which can be formulated as
\begin{align}
    H^t &= transformer^t(X^t), \\
    h^t &= cls(H^t), \\
    S^t(s) &= norm(h^t),
\end{align}
where $X^t$ is the feature embedding of item's text features like title, category, brand, {\etc}, $transformer^q$ is a $o^t$-heads, $l^t$-layers transformer encoder. Here use $o^t=12$ and $l^t=1$.

Similarly, we can get the output of \textbf{Item Image Tower} as $S^i(s)$, and the output of \textbf{Item Multimodal Tower} as $S^m(s)$. What's different in item multimodal tower is that the transformer layer receives three inputs as $Q$, $K$, $V$, denoted as
\begin{align}
    H^m &= transformer^m(H^t, H^i, H^i).
\end{align}
This design aims to conduct better modality fusion through the cross attention in $H^m$.

\subsection{Multi-task Learning-based Pre-training Model}
\label{sec:mlmm}
Modality fusion and alignment are two major challenges in our work. In order to achieve better performance, we design a multi-task learning-based pre-training model that contains three tasks as shown in the left part of Fig.~\ref{fig:three_loss}. Respectively, the first task aims to learn semantic projection of text modality between query and item, the second task aims to learn modality alignment between query's text  and item's image, and the last one aims to conduct the asymmetric modality alignment between query and item.


\subsubsection{Model Architecture.} \label{sssec:model_arch} In detail, for the first semantic projection task, we compute the score for query $q$ and item $s$ as
\begin{equation}
    f^t(q, s) = F(Q(q), S^t(s)),
\end{equation}
where $f^t(q, s)$ dedicates to measure the semantic similarity of text modality between query and item. Similarly, we can get the score for second modality alignment task as $f^i(q, s)$, which denotes the alignment performance of query's text and item's image, and the score for the last task as $f^m(q, s)$, which denotes the alignment performance of query's text and item's multi-modality.


\subsubsection{Loss Function.} In order to further improve the modality alignment and fusion, we also enrich the training dataset by a sampling strategy proposed in~\cite{qiu2022pre}. 
As a result, the number of synthetic queries is more than 100 times of true queries. For negative examples, we conduct batch negative sampling strategy as that in work~\cite{zhang2020towards}.
Then the pre-training dataset can be formulated as
\begin{equation}
\small D = \left\{\left(q_i, s^+_i, N_i\right) \; \left| \; i,\; r(q_i, s^+_i) = 1, \; r(q_i, s^-_j) = 0\; \forall\; s^-_j \in N_i \right\}\right.,
\end{equation}
where the triplet $(q_i, s^+_i, N_i)$ denotes a training example, which contains a synthetic query $q_i$, a positive item $s^+_i$ that is relevant to the query and the relation is denoted as $r(q_i, s^+_i) = 1$, a negative item set $N_i$ that consists of the sampled negative items which are irrelevant to the query and the relation is denoted as $r(q_i, s^-_j) = 0$. Now we employ cross entropy loss over the training dataset as
\begin{align}
    \label{eq:softmax}
    p &= softmax(\{f(q_i, s_i) \; | \; s_i \in \{s_i^+\}\cup{N_i}\}), \\
    \small \mathcal{L}(D) &= \sum_{(q_i, s^+_i, N_i) \in D} \sum_{s_j \in \{s^+_j\}\cup N_j}r(q_j, s_j) \; log(p_j).
\end{align}

For the three tasks in~\ref{sssec:model_arch}, we apply each scoring function in Formula~\ref{eq:softmax} and denote the loss as $\mathcal{L}^t(D)$, $\mathcal{L}^i(D)$ and $\mathcal{L}^m(D)$. 
Considering the different contributions of text and image, we apply different weights to each loss, thus the final loss is computed as
\begin{equation}
    \mathcal{L}(D) = {\alpha}\mathcal{L}^t(D) + {\beta}\mathcal{L}^i(D) + {\gamma}\mathcal{L}^m(D),
    \label{eq:loss}
\end{equation}
where $\alpha$, $\beta$ and $\gamma$ are hyperparameters.

\subsection{Adaptive Embedding Learning-based Fine-tuning Model}
\label{sec:aelmm}
In online e-commerce system, we find that image benefits to item retrieval with different importance for different product categories. For example, image color is very important in fashion categories, like clothes, bags, shoes, {\etc}, to help retrieve items relevant to query within color words, while it is less important for categories like books or medicine where color is not a key attribute. To avoid introducing abundant information for categories which are not in need of image, we propose an adaptive embedding learning-based multimodal model for the fine-tuning stage.

\subsubsection{Model Architecture.}
The right part of Fig.~\ref{fig:three_loss} illustrates the architecture of our fine-tuning model, which contains three towers, {\ie} query tower $Q$, item text tower $S^t$ and item multimodal tower $S^m$ that share the same structure and parameters with our pre-training model.
Moreover, we use a prediction header $P$ to decide whether introducing image into item representation.
$P$ is a jointly learned task with an extra optimizer supervised by a fashion dataset.
When $P(q)=1$, we introduce image modality to learn item multimodal embedding, otherwise, learn item text embedding.
As a result, image contributes to item representation dynamically through adaptive item embedding decided by the result of query prediction header.

\subsubsection{Loss Function.} In e-commerce system, user click logs, which show implicit relevance between query and item, have been proven useful to realize semantic retrieval~\cite{zhang2020towards,zhang2021joint,xia2021searchgcn,jiang2022givens}. Here we also explore click logs to train our fine-tuning model.
Similar with the pre-training model, we employ cross entropy loss over the training dataset from click logs, and denote the loss with scoring function $f^m$ as $\mathcal{L}^{\prime m}$, denote the loss with scoring function $f^t$ as $\mathcal{L}^{\prime t}$. Then the final loss can be formulated as
\begin{align}
    \mathcal{L}^{\prime}(D) = P\mathcal{L}^{\prime m}(D) + (1-P)\mathcal{L}^{\prime t}(D).
\end{align}


\section{Experiments}
\label{sec:exp}
In this section, we first introduce our industrial datasets, model metrics and experimental setup, then demonstrate the experimental results of our proposed model along with offline evaluations and online A/B tests.

\begin{figure*}[th!]
    \centering
    \begin{subfigure}[b]{0.3\textwidth}
        \centering
        \includegraphics[width=4cm]{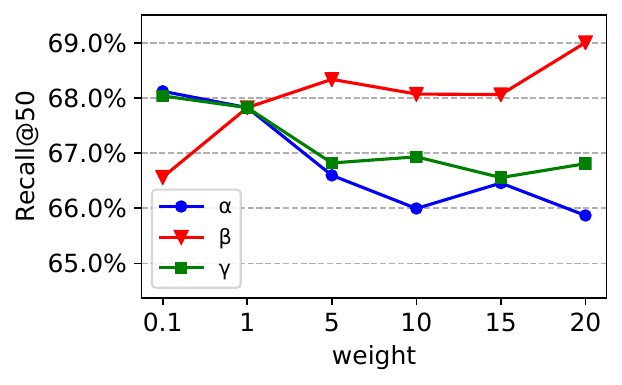}
        \caption{Overall Dataset}
        \label{fig:compare_all}
    \end{subfigure}
    \begin{subfigure}[b]{0.3\textwidth}
        \centering
        \includegraphics[width=4cm]{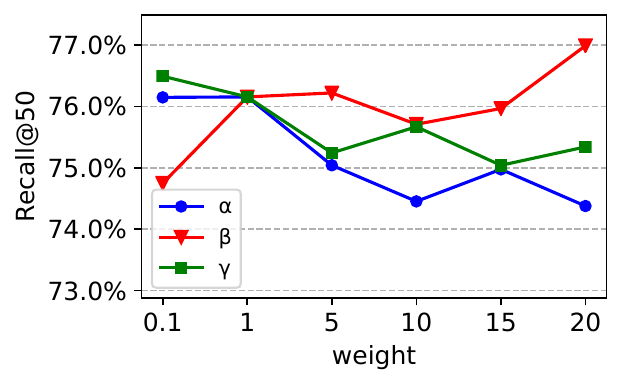}
        \caption{Fashion Dataset}
        \label{fig:compare_color}
    \end{subfigure}
    \begin{subfigure}[b]{0.3\textwidth}
        \centering
        \includegraphics[width=4cm]{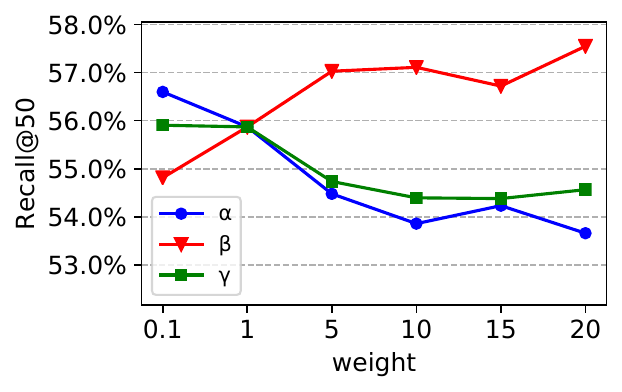}
        \caption{Not-fashion Dataset}
        \label{fig:compare_uncolor}
    \end{subfigure}
    \caption{Effect of the weights $\alpha$, $\beta$, $\gamma$ for loss functions on overall dataset, fashion dataset and not-fashion dataset. For each line in the figures, we adjust one parameter(for example $\alpha$) and set another two to 1 (for example $\beta=1$ and $\gamma=1$).}
    \label{fig:mod_contri}
\end{figure*}

\begin{table*}[th!]
\centering
\caption{Comparision between the baseline methods and our proposed models.}
\begin{tabular}{c|ccc|ccc|ccc}
\hline
       & \multicolumn{3}{c|}{Overall} & \multicolumn{3}{c|}{Fashion} & \multicolumn{3}{c}{Not-fashion} \\ \cline{2-10} 
       & R@50     & P@50     & F1@50     & R@50     & P@50     & F1@50     & R@50      & P@50      & F1@50      \\ \hline\hline
DPSR-i &  0.349  &  0.018  &  0.034  &  0.485  &  0.022  & 0.042   &  0.154  &  0.011  &  0.021 \\
DPSR   &  0.641  &  0.031  & 0.059   &  0.749  &  \textbf{0.032}  &  \textbf{0.061}  &  0.486  &  0.030  & 0.057  \\
SMAR-ni   &  0.663  &  0.032  & 0.061   &  0.746  & 0.031   &  0.060  &  0.544  &  0.033  & 0.062  \\
SMAR-nt   &  0.671  &  0.032  &  0.061  &  0.754  &  0.031  &  0.060  &  0.551  &  0.033  &  0.062 \\
SMAR-nm   &  0.681  &  \textbf{0.033}  &  \textbf{0.063}  &  0.759  &  \textbf{0.032}  &  \textbf{0.061}  &  0.569  &  \textbf{0.034}  &  \textbf{0.064} \\
SMAR-nc   &  0.670  &  0.032  &  0.061  &  0.755  & 0.031   &  0.060  &  0.548  &  0.033  &  0.062 \\
SMAR  &  \textbf{0.690}  &  \textbf{0.033}  &  \textbf{0.063}  &  \textbf{0.770}  &  \textbf{0.032}  & \textbf{0.061}   & \textbf{0.575}   &  \textbf{0.034} & \textbf{0.064}  \\ \hline
\end{tabular}
\vspace{-2mm}
\label{tab:exp_results}
\end{table*}


\subsection{Experimental Setup}
\textbf{Datasets.} The statistics of our dataset, which is sampled from 60 days user click logs, is shown in Table~\ref{tab:dataset}. The dataset is further divided into fashion and not-fashion datasets by the truth that whether the item belongs to fashion categories like clothes, shoes, bags, {\etc} or not. We have open source the overall dataset for the sake of the reproduction of our work and further public researches.

\begin{table}[h!]
\centering
\caption{Dataset statistics.}
\small
\begin{tabular}{c|ccc}
\hline
Dataset   & \# Examples &  \# Queries & \# Items  \\ \hline\hline
Overall   & 2,163,445   &  373,343  &  957,842 \\ 
Fashion  &  1,035,587  & 213,708  &  298,906   \\ 
Not-fashion  & 1,127,858 &  159,635 & 736,971\\ \hline
\end{tabular}
\vspace{-2mm}
\label{tab:dataset}
\end{table}

\textbf{Metrics.} We evaluate our models by metrics of recall@$k$(R@$k$), precision@$k$(P@$k$) and F1-score@$k$(F1@$k$), which are standard retrieval quality metrics used in many works~\cite{zhu2018learning,zhang2020towards}.


\textbf{Setup.} Both pre-training and fine-tuning models are trained with 2 Tesla A100 GPU cards, and optimized by AdamW algorithm~\cite{loshchilov2017decoupled} with learning rate 0.00005 and batch size 128 per gpu.

\subsection{Experimental Results}
\subsubsection{Model Evaluations.}
As shown in table~\ref{tab:exp_results}, we compare the performance between several baselines and our proposed models.
\begin{itemize}
    \item DPSR refers to the semantic retrieval model for e-commerce system proposed in~\cite{zhang2020towards}, which is a strong baseline that has been successfully deployed in a popular retail platform.
    \item DPSR-i refers to a variant of DPSR which uses image modality instead of text modality for item representation.
    \item {\smar} refers to our proposed semantic-enhanced modality-asymmetric retrieval model.
    \item {\smar}-ni, {\smar}-nt and {\smar}-nm refer to the variants of {\smar} which have no modality alignment task, semantic projection task, modality fusion and asymmetric modality alignment task in pre-training model respectively.
    \item {\smar}-nc refers to a variant of {\smar} which has no cross attention mechanism in the item multimodal tower.
\end{itemize}

In order to explore more detailed performance of our models, we evaluate our models not only on the overall dataset, but also on fashion dataset and not-fashion dataset. From the comparison results in Table~\ref{tab:exp_results}, we can make the following observations:
\begin{itemize}
    \item Our proposed model {\smar} outperforms baseline models DPSR and DPSR-i significantly on all datasets, with 4.90\% and 1.50\% improvements for R@50 and P@50 respectively on overall dataset, which indicates that {\smar} is quite effective to enhance semantic retrieval by introducing image modality. 
    \item Comparing the results of {\smar} and the variant models {\smar}-ni, {\smar}-nt, {\smar}-nm, we can see that the multi-task learning objective described in Section~\ref{sec:mlmm} performs well in asymmetric modality alignment.
    \item Comparing the results of {\smar} and {\smar}-nc, we can see that the cross attention between text and image is a successful mechanism to tackle the challenge of modality fusion, which captures supplementary information from image.
\end{itemize}

\subsection{Modality Contributions}
The performance of DPSR and DPSR-i shows that text and image have different contributions to the latent representation of item. Thus we apply modality weights in Equation~\ref{eq:loss} to facilitate modality contribution conditioning to achieve better model performance.

As shown in Fig.~\ref{fig:mod_contri}, the values of hyperparameters $\alpha$, $\beta$, $\gamma$, which indicate the weights for text modality, image modality and multimodality respectively, vary from 0.1 to 20 to explore the modality contributions.
According to the results, we can see that (1) modality weights have obvious effect on model performance, each modality has different contribution. (2)$\alpha$ and $\gamma$ have similar trends, which may due to the truth that text provides more information than image which have been proven in the experimental results in above section, thus they are more relevant to each other. 

\subsection{Online A/B Tests}

We conducted live experiments over 15\% of the entire traffic using a standard A/B testing configuration in our online experiments platform which is built based on the theory proposed in~\cite{tang2010overlapping}. In the test group, our model {\smar} retrieves another 1000 candidates in addition to the results in baseline group.
Moreover, our e-commerce platform is a mature system which also contains query parsing, item relevance controlling and item ranking modules, which have been tuned by hundreds of researchers and engineers for years, thus we declare that the current system is a very strong baseline.

The results of online A/B tests are shown in Table~\ref{tab:online_exp}. We focus on two main core business metrics, gross merchandise value (GMV) and user conversation rate (UCVR), which are regarded as standard indicators for business gains. In consideration of confidential business information protection, we only present relative improvements. We can observe that our proposed model {\smar} achieves improvements in online e-commerce system, especially in fashion categories, which proves {\smar}'s success in business scenario.

\begin{table}[h!]
    \vspace{-3mm}
    \caption{Online A/B test improvements.}
    \centering
    \begin{tabular}{c|r r r } \hline
            & GMV  & UCVR  \\ \hline\hline
    {\smar}  & +0.285\% & +0.174\% \\
    {\smar} in fashion categories  & +1.112\% & +0.437\% \\
    \hline 
    \end{tabular}
    \vspace{-5mm}
    \label{tab:online_exp}
\end{table}


\section{Conclusion}
In this paper, we have proposed a novel semantic-enhanced modality-asymmetric retrieval model (SMAR), which adopts a two stage training strategy and has been deployed on an online e-commerce search system to serve the main traffic. SMAR formulates pretraining as a multi-task learning problem while finetuning as an adaptive embedding learning problem. Results on both offline experiments and online A/B tests show that the proposed model performs significantly better than unimodal models in search system.

\bibliographystyle{ACM-Reference-Format}
\balance

\bibliography{references}
\appendix
\section{Presenter Profiles}
\header{ZhiGong Zhou} is a researcher in the Department of Search and Recommendation at JD.com. He received his master degree from Institute of Automation, Chinese Academy of Sciences. His research focuses on information retrieval.

\header{Ning Ding} is a researcher in the Department of Search and Recommendation at JD.com. She received her master degree from South China University of Technology. Her research focuses on information retrieval and vision-language modeling.

\header{Han Zhang} is a senior researcher in the Department of Search and Recommendation at JD.com. She had about 5 years experience in search system research.
She received her master degree from Natural Language Processing Lab, Department of Computer Science and Technology, Tsinghua University.

\end{document}